# Towards Distributed Convoy Pattern Mining

Faisal Orakzai[†], Thomas Devogele[‡], Toon Calders[†]

[†]Department of Computer & Decision Engineering (CoDE) Université Libre de Bruxelles 1050 Bruxelles, Belgium
[‡]Department of Informatics Université Francois Rabelais 41000 BLOIS, France
[†]{ofaisal,tcalders}@ulb.ac.be ,[‡]thomas.devogele@univ-tours.fr

## ABSTRACT

Mining movement data to reveal interesting behavioral patterns has gained attention in recent years. One such pattern is the convoy pattern which consists of at least $m$ objects moving together for at least $k$ consecutive time instants where $m$ and $k$ are user-defined parameters. Existing algorithms for detecting convoy patterns, however do not scale to real-life dataset sizes. Therefore a distributed algorithm for convoy mining is inevitable. In this paper, we discuss the problem of convoy mining and analyze different data partitioning strategies to pave the way for a generic distributed convoy pattern mining algorithm.

**Categories and Subject Descriptors:** H.2.8 Database Applications: Data mining

## 1. INTRODUCTION

Convoys [10] are found by performing density based clustering such as by DBSCAN [4] on object locations at each time instant followed by combining these clusters over the time dimension. Finding convoy patterns can be useful in many application domains. It can be used to find groups of people traveling together by public transport or to determine potential candidates for carpooling. For mining convoy patterns, various algorithms [10, 11, 16, 1] have been proposed, however the existing algorithms have been tested only on small datasets which can easily fit into memory. The most efficient existing convoy pattern mining algorithm [11] took 100 seconds during an experimental run on a small dataset containing a couple of hours of movement data of only 13 cattle hence a distributed and scalable algorithm for convoy mining is necessary. A vital part of distributed algorithms is their data partitioning strategy which heavily influences their performance as well as accuracy. In this paper, we elaborate desired data partitioning properties and analyze different data partitioning strategies for mining convoy patterns in a distributed shared-nothing architecture based on these properties.

## 2. RELATED WORK

The algorithms *CuTS, CuTS+* and *CuTS\** [11] try to reduce the cost of DBSCAN by reducing the number of objects $n$ in each run using a filter-and-refine paradigm. In the first phase, each object's trajectory is simplified by reducing the number of time-location pairs using Douglas-Peucker algorithm (DP) [3]. The simplified trajectories are then partitioned into pieces each corresponding to a time duration $\lambda$. For each time duration $\lambda$, the pieces(sub-trajectories) are clustered using the DBSCAN algorithm. This step reduces the dataset to only those object trajectories which have the potential to form convoys. In the second step, the *CMC* [11] algorithm is applied on the reduced dataset. Although *CuTs* family of algorithms has proven faster than other methods, the cost of trajectory simplification is $\mathcal{O}(T^2 N_t)$ where $T$ is the number of points in a trajectory and $N_t$ is the number of trajectories in the dataset. Using trajectory simplification also disallows us from using the same indexing structure as that of DBSCAN. The index for DBSCAN is based on location whereas the index required for trajectory simplification is based on object identity. Finding the right combination of parameters for CuTs family, for it to give acceptable performance is very hard and may involve multiple expensive iterations with no guarantee of success.

In [16], Yoon et al. discovered that the convoy mining algorithms proposed by Jeung et al. [11] have serious problems with respect to accuracy and recall. They refer to the convoy pattern proposed by Jeung et al. as *Partially Connected Convoy* and propose a contextually different version of the convoy pattern called *Valid Convoy*. They present a corrected version of *CMC* algorithm called *PCCD*. As CuTs family of algorithms is based on CMC, they also have serious accuracy issues.

## 3. THE CONVOY PATTERN

The convoy pattern is based on density conection which can be defined as follows:

*Definition 1.* **(Density-Connected)** *Given a set of points $S$, a point $p \in S$ is* density-connected *to a point $q \in S$ with respect to $\epsilon$ and $m$ if there exists a point $x \in S$ such that both $p$ and $q$ are density-reachable from $x$.*

*Definition 2.* **(Snapshot Cluster)** *Given an integer $m$ and distance threshold $\xi$, a set $c$ of density connected points at time instant $t$ w.r.t $\xi$ is called a* snapshot cluster *if $|c| \geq m$, all $p \in c$ are density connected and $\nexists r \notin c$ that is density connected to any point in $c$.*

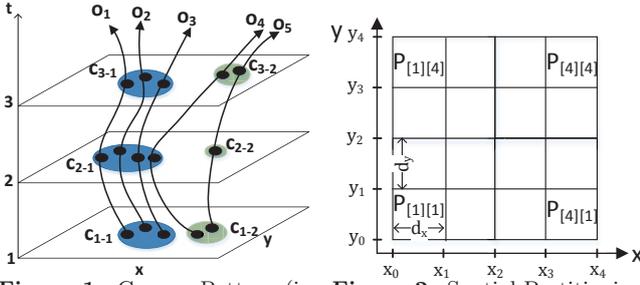

**Figure 1:** Convoy Pattern (in blue) with $m = 3$ and $k = 3$

**Figure 2:** Spatial Partitioning

In Figure 1, for $m = 3$, clusters $c_{1-1}$, $c_{2-1}$ and $c_{3-1}$ are snapshot clusters as they are sets of density connected points with $size \geq m$. A *trajectory* is the movement trace of an object and contains all the points $(x, y)$ traversed by the object in the order of traversal. Each point has a time $t$ associated with it. In other words, a trajectory can be described as a temporal version of a line. A snapshot clusters at a time $t_x$ can be found by querying all tuples with $t = t_x$ to retrieve all objects present at time $t_x$ and their locations, and performing density based clustering. Jeung et al. define a convoy query as follows:

*Definition 3. (Convoy Query)* Given a set of trajectories of $N$ objects, a distance threshold $\xi$, an integer $m$, and an integer lifetime $k$, the convoy query returns all possible groups of objects, so that each group consists of a (maximal) set of density-connected objects with respect to $\xi$ and $m$ during at least $k$ consecutive time points.

Using the above definition, a convoy can be defined as:

*Definition 4. Convoy* Given a set of objects $O$, a convoy $v$ is a pair $(O, [t_s, t_e])$ such that $|O| \geq m$, $t_e - t_s + 1 \geq k$ and for all $t_s \leq t \leq t_e$, $O$ is density connected at time $t$. Further more, $v$ cannot be further extended, that is: for all $x \notin O, (O \cup \{x\}, [t_s, t_e]), (O, [t_s - 1, t_e]), (O, [t_s, t_e + 1])$ do not satisfy this constraint.

In Figure 1, the set of objects $O = \{o_1, o_2, o_3\}$ forms a convoy with $m = 3$ and $k = 3$ which can be represented as $(O, [1, 3])$. To define a convoy in terms of snapshot clusters, we can say that it is a set of at least $m$ objects present in the same snapshot cluster for $k$ consecutive timestamps. A single snapshot cluster can be considered as a convoy with $k = 1$.

## 4. PROBLEM OVERVIEW

A distributed algorithm for convoy pattern mining can be divided into three stages: partitioning, local convoy pattern minings and merging to produce the global result. The primary step in the development of a distributed convoy mining algorithms is to find an efficient data partitioning strategy with the following properties:

- **Data Exchange:** As the cost of communication between the nodes of a distributed system is high, less data should be exchanged between the nodes of the cluster. This can be done by locating the data processed together at the same node, also called $co-location$
- **Data Redundancy:** Data located on different nodes should be less redundant. More redundancy means more CPU and disk space consumption which leads to slower processing speeds.
- **Partitioning Costs:** The parameters of the partitioning strategy should be simple to decide and shouldn't depend on data statistics. It is not realistic to build statistics on huge datasets before partitioning because of high costs involved.
- **Disk Seeks:** The data within a partition should also follow the co-location principle to minimize disk seeks.
- **Data Ordering:** The data inside the partition should be stored in the order of processing. This property omits the requirement of an index and allows streaming algorithms to operate on the data. Building an index on huge datasets is not feasible. Data ordering allows faster disk access as the operating system can perform efficient paging.

Our target dataset has 4 columns with the schema $< oid, x, y, t >$ where $oid$ is the object id, $x$ and $y$ represent the location in two dimensions, and $t$ is the time instant at which the the object was at that location. Because of its scalability, we only consider data parallelism strategy. The data parallelism strategy involves the division of data among multiple processors/nodes and the execution of an identical algorithm on each of the nodes. Trajectory data can be partitioned on moving object properties or on the spatial or temporal dimension. The choice of the partitioning strategy heavily influences the accuracy and performance of a distributed algorithm. In the following, we analyze different partitioning strategies for distributed convoy mining.

### 4.1 Object Based Partitioning

We can partition the trajectory data based on object properties e.g. oid. As convoy mining involves clustering of objects on their spatial dimension, object based partitioning violates the *colocation* principle hence resulting in heavy data exchange. The second stage of a convoy mining algorithm is to combine the spatial clusters on temporal dimension. As both the phases of convoy mining (clustering and combining) do not involve object dimension, there is no acceptable data redundancy level that can achieve acceptable data exchange.

### 4.2 Spatial Partitioning

A simple spatial partitioning strategy is to divide the data in equal pair-wise disjoint partitions over $x$ and $y$ dimensions. Thus, the complete dataset can be seen as a two dimensional array $P_{[i][j]}$ where $i = 1 \ldots n_x$ and $j = 1 \ldots n_y$ such that $\cap_1^n P_{[i][j]} = \phi$. Let $n_p = n_x * n_y$ be the number of total partitions and $N$ be the number of available nodes in our target cluster. If $n_p > N$, some nodes will have to host more than one partition. For illustration purposes, for the rest of this section we assume that $n_p = N$ thus each node contains only one partition. Figure 2 shows an example of the above spatial partitioning strategy. Other complex partitioning strategies exist such as Partitioning with Reduced Boundary Points (PRBP) [8] and the cost based partitioning for skewed data [7]. However we will see later that spatial partitioning is not ideal for solving distributed convoy mining problem. At the conceptual level, convoy mining consists of two phases, density based clustering and formation of convoys based on the detected clusters. Spatial partitioning brings problems in both phases which we explain in the following with the help of Figure 3. The figure shows three adjacent partitions incrementing in $x$ dimension. As assumed before, each of these partitions resides

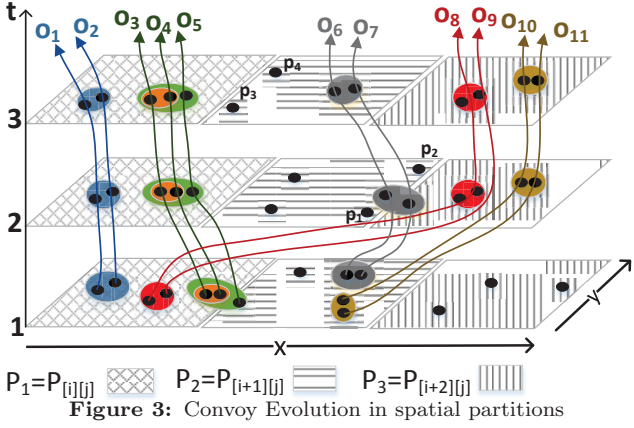

P₁=P_{[i][j]}  P₂=P_{[i+1][j]}  P₃=P_{[i+2][j]}
**Figure 3:** Convoy Evolution in spatial partitions

| Table 1: Partitioning Strategies Comparison |||| 
|---|---|---|---|
| Properties | Object | Spatial | Temporal |
| Data Exchange | Too High | Too High | Low |
| Data Redundancy | variable | High | No |
| Partitioning Cost | $O(n)$ | $O(n)$ | No |
| Disk Seeks | High | High | No |
| Data Ordering | No | No | Yes |

on a different node. For the purpose of illustration, we call these partitions $P_1$, $P_2$ and $P_3$. The y-axis shows the data contained by each partition corresponding to time. Black ellipses show object locations whereas colored ellipses represent snapshot clusters. At a particular time instant, objects are represented by their locations, we use the term '*object*' instead of the term *location* or *point* for better intuition. For this example, we assume $m = 2$ and $k = 3$ for convoy pattern mining. For mining convoy patterns, we must first mine convoy patterns locally on each partition and then combine the results at a central node. The blue convoy represents an ideal situation in which all object locations/points belonging to a snapshot cluster exist in one partition hence this convoy will correctly be detected by local convoy pattern mining. In the following we discuss potential problems in spatial partitioning strategy.

*Partial or Missing Clusters*
During spatial partitioning, the objects of a potential snapshot cluster may get split in different partitions. If in a partition, the number of objects belonging to the split cluster is equal to or greater than $m$, a partial cluster will be detected. For instance in the figure 3, density based clustering on partition $P_1$ at time $t = 1$, detects a smaller orange cluster instead of the green cluster. Resultantly, a smaller orange convoy with objects $o_{[3,4]}$ is detected instead of the green convoy with objects $o_{[3,5]}$. Similarly, If the number of the split cluster's objects present in the current partition is less than $m$, the cluster will not be detected at all. In the figure, the gray cluster at time $t = 2$ is split between partitions $P_2$ and $P_3$ such that it's size in each partition is less than $m$. Thus, the gray cluster won't be detected in either partitions, thus missing the gray convoy. This problem can be solved by:

1. *Overlapped Partitions:* Instead of having disjoint partitions, data can be partitioned in such a way that the partitions overlap at their edges. Overlapped partitions can solve the problem of partial/missing clusters but causes redundancy, hence extra storage cost. Additionally, it is not easy to determine the right overlapping factor. If the cluster sizes are larger than the estimated overlap, the clusters will not be detected accurately however if the overlap is much larger than cluster sizes, a lot of storage capacity will be wasted. As cluster sizes are not known in advance, we can scan the data to build statistics for cluster sizes but that raises partitioning cost.
2. *Reducing $m$*: Another possible way to solve the problem of partial/missing clusters could be to reduce $m$ for the clusters near the partition edge. The safest choice for a reduced $m$ is 1 which will cause the clustering algorithm to consider all noise points as clusters thus increasing data exchange. Additionally, a complex procedure will be required to merge partial clusters into full clusters.

*Partial Convoys*
The second phase of convoy pattern mining is to use the clusters to find out convoys. In this phase, we assume that the problem of cluster split has already been handled by the use of appropriate strategy and we have found accurate clusters in each partition. In partition $P_1$, we can combine all blue clusters and green clusters to form a blue and a green convoy respectively. Similarly, in partition $P_2$ we can combine all gray clusters to form the gray convoy. In partition $P_3$, we can not find any convoy because combining red and brown clusters result into convoys with the lifetime of 2 ($T = [3, 4]$) which is less than the desired parameter $k = 3$. The reason for the failure in detecting the brown convoy is that it is split between partitions $P_2$ and $P_3$ and the convoy pattern mining algorithm running on partition $P_3$ is unaware of the existence of the brown cluster at time $t = 1$ in partition $P_2$. The solution to this problem could be to share all the clusters between adjacent partitions so that the local convoy pattern mining algorithm can detect convoys accurately. There can be millions of clusters in a partition which are not part of any convoy causing high data exchange.

*Fast Moving Objects*
Another problem with convoys spanning multiple partitions is their speed. If the size of the partitions is small and the speed of a convoy is very fast, it is possible that it may not appear in adjacent partitions. In the figure, red convoy is a fast moving convoy. At time $t = 1$, it is present in partition $P_1$ but at time $t = 2$, it appears in partition $P_3$ instead of partition $P_2$. In this case, the algorithm running on partition $P_3$ won't be able to detect the red convoy unless it is aware of the presence of the red cluster in partition $P_1$ at time $t = 1$. Sharing clusters between adjacent partitions can not solve this problem. We must share the clusters between non-adjacent partitions as well which is highly network intensive. The worst case cost for this approach is $O(P^2)$ which means that the clusters from each partition will have to be merged with every other partition to achieve correct results. This is certainly a no-go area as it causes unacceptably high data exchange.

*Cost of DBSCAN*
Spatial partitioning performs a bit better in terms of cost of DBSCAN execution. The cost of performing density based clustering using DBSCAN on spatial partitions at time $t$ is $O(\frac{n}{N} \log \frac{n}{N} * N)$ where $n$ is the total number of objects present at time $t$ and $N$ is the number of partitions. The DBSCAN cost for clustering objects present at a time-stamp using temporal partitioning (as we will see later) is $O(n \log n)$ which is higher when compared to spatial partitioning.

## 4.3 Temporal Partitioning

In contrast to spatial partitioning, temporal partitioning does not pose above mentioned problems in convoy pattern mining, the reason being the restriction in the movement of objects in the temporal dimension. As the temporal dimension is unidirectional, an object can only move forward in the temporal direction. Thus, the convoy mining results from a partition only needs to be merged with the results from the next or previous partition. This significantly reduces the size of the data to be transported over network for merging. In this section we propose a distributed and disjoint time coherent partitioning strategy based on which we can model the distributed convoy pattern mining problem.

Let $DB$ be the complete set of movement data, $N$ be the number of computational nodes available. Let $t_s(DB) = t_0 < t_1 < t_2 \cdots < t_{N-1} < t_N = t_e(DB)$. The time coherent partitioning of $DB$ based on the split points $t_0, \ldots, t_N$ is defined as: $DB = P_1 \cup \cdots \cup P_N$ where: $P_i = \{(oid, x, y, t) \in DB | t_{i-1} < t \leq t_i\}$. Here, $t_s(P_i)$ and $t_e(P_i)$ denote the start and end time of the partition $P_i$. Notice that:

$$t_s(P_i) = t_{i-1} + 1 \qquad t_e(P_i) = t_i$$
$$t_s(P_1) = t_s(DB) \qquad t_e(P_N) = t_e(DB)$$

The timespan $T_i$ of partition $P_i$ for $i = 1 \ldots N$ is defined as:

$$T_i = [t_s(P_i), t_e(P_i)]$$

As in an embarrassingly parallel job, the total execution time is determined by the execution time of the slowest node [15], data should be equally distributed in all partitions so that any one partition does not take significantly longer time to process and hence act as a bottleneck. For uniformly distributed data over time, partitions with equal number of time-stamps can be created with timespan:

$$T_i = \left[ \left\lceil \frac{|T_{DB}|}{N} * (i-1) \right\rceil + 1, \left\lceil \frac{|T_{DB}|}{N} * i \right\rceil \right]$$

For skewed datasets with imbalanced distribution of data over time, equal partitions over time is suboptimal. For such datasets, different partitioning strategies have been proposed in [2, 9, 13]. Some distributed data processing platforms implement their own partitioning strategy e.g. MapReduce [5] partitions the data based on HDFS block size which ensures equal data size for each of its processing unit (Mapper).

Table.1 shows the comparison between the partitioning strategies. Temporal partitioning ensures no data exchange for local convoy pattern mining where as low data exchange in the global merging phase. As the data is temporally sorted naturally, there are no partitioning costs. The blocks in a distributed file system such as HDFS can be taken as partitions. Temporal partitioning also follows the colocation principle in both the phases of convoy pattern mining i.e. clustering (as all the locations for a timestamp exist on one node) and merging (as data is naturally sorted temporally and temporally adjacent clusters are required to be merged).

## 5. CONCLUSION

Convoy pattern mining is computationally expensive and existing algorithms do not scale up to the huge amounts of movement data. In this paper we define desirable properties for a partitioning strategy and analyze differing partitioning strategies based on those properties. We found temporal partitioning more suitable to distributed convoy mining problem because of its merits and natural fit for storage on HDFS.

**Acknowledgment.** This research was partially funded by "The Erasmus Mundus Joint Doctorate in Information Technologies for Business Intelligence - Doctoral College (IT4BI-DC)".